\documentclass[aps,prb,twocolumn,showpacs,groupedaddress]{revtex4}
\usepackage{graphicx}

\begin{document}

\title{Pressure-Temperature Phase Diagram of Multiferroic $Ni_3V_2O_8$}
\author{R. P. Chaudhury$^{1}$, F. Yen$^{1}$, C. R. dela Cruz$^{1}$, B. Lorenz$^{1}$, Y. Q. Wang$^{1}$, Y. Y. Sun$^{1}$, and C. W. Chu$^{1,2,3}$}
\affiliation{$^{1}$TCSUH and Department of Physics, University of
Houston, Houston, TX 77204-5002} \affiliation{$^{2}$Lawrence
Berkeley National Laboratory, 1 Cyclotron Road, Berkeley, CA 94720}
\affiliation{$^{3}$Hong Kong University of Science and Technology,
Hong Kong, China}
\date{\today }

\begin{abstract}
The pressure-temperature phase diagram of multiferroic $Ni_3V_2O_8$
is investigated for hydrostatic pressures up to 2 GPa. The stability
range of the ferroelectric phase associated with the incommensurate
helical spin order is reduced by pressure and ferroelectricity is
completely suppressed at the critical pressure of 1.64 GPa at 6.2 K.
Thermal expansion measurements at ambient pressure show strong
step-like anomalies of the lattice parameters associated with the
lock-in transition into the commensurate paraelectric phase. The
expansion anomalies are highly anisotropic, the related volume
change is consistent with the high-pressure phase diagram.
\end{abstract}

\pacs{75.30.-m,75.30.Kz,75.50.Ee,77.80.-e,77.84.Bw} \maketitle











The Kagom\'{e} staircase compound $Ni_3V_2O_8$ is one of the
multiferroic magnetoelectric compounds that has recently attracted
attention because of the complex phase sequence of several magnetic
transitions upon decreasing temperature and the appearance of
ferroelectricity in one of the magnetic
phases.\cite{lawes:04,lawes:05} The origin of ferroelectricity in a
limited temperature range (3.9 K $<$ $T$ $<$ 6.5 K) has been a
matter of discussion and it was associated with a particular type of
magnetic order of the nickel spins. The nickel ions in this compound
form a quasi-planar buckled Kagom\'{e} staircase in which magnetic
frustration is immanent due to the geometry and the
antiferromagnetic (AFM) exchange interactions. Due to the buckling
of the Kagom\'{e} plane there are inequivalent nickel sites and the
magnetic exchange interactions between the different nickel spins
include first and second nearest-neighbor isotropic Heisenberg
interactions as well as magnetic anisotropy and the antisymmetric
Dzyaloshinskii-Moriya interactions.\cite{lawes:05} This multitude of
magnetic interactions and the geometrical frustration in the
Kagom\'{e} structure give rise to a cascade of magnetic phase
transitions upon decreasing temperature. The magnetic structure of
$Ni_3V_2O_8$ has been completely resolved by neutron scattering
experiments only recently.\cite{lawes:04,kenzelmann:06} The
high-temperature paramagnetic phase is orthorhombic, space group
Cmca (No 64). With decreasing temperature, magnetic order sets in at
$T_{N}$=9.8 K into an incommensurate (IC) sinusoidal (collinear)
spin structure (HTI phase, we follow the notations of Lawes et
al.\cite{lawes:04}). At $T_{C1}$=6.5 K the spin structure changes to
a helical spin density wave breaking the spatial inversion symmetry
with the onset of ferroelectricity (LTI phase). Upon further
cooling, at $T_{C2}$=3.9 K, a phase transition into a commensurate
magnetic structure takes place (C phase). The C phase is
paraelectric and the magnetic order in this phase is symmetric with
respect to the spatial inversion operation. The magnetic symmetries
in the HTI, LTI, and C phases have been discussed in great detail in
recent publications.\cite{harris:06,harris:06b,kenzelmann:06}
Another transition into a second commensurate magnetic phase at
$T_{CC'}$ (C' phase) concludes the sequence of transitions. The
magnetic phase diagram has been explored
extensively.\cite{lawes:04,lawes:05,kenzelmann:06,wilson:06}

The improper ferroelectricity observed in the LTI phase is induced
as a secondary order by the helical spin density wave via the
spin-lattice coupling. Various models based on symmetry properties
of the crystal and the magnetic order have been
discussed.\cite{lawes:05,harris:06b} A ferroelectric (FE)
polarization can arise from a third order coupling of two
non-collinear components of a magnetic order parameter and the
polarization.\cite{lawes:05,mostovoy:06} A helical spin density wave
as observed in $Ni_3V_2O_8$ in the LTI phase fulfills the condition
for this coupling between the polarization and the magnetic order
parameters and it breaks the inversion symmetry of the magnetic
structure. A strong spin-lattice interaction couples the lattice to
the magnetic order and results in the local displacements with a
macroscopic polarization. A microscopic model describing the
magnetoelectric coupling in $Ni_3V_2O_8$ was recently developed by
Harris et al.\cite{harris:06} showing that very small atomic
displacements can give rise to the observed FE distortions.
Therefore, tuning the lattice displacements by controlled
introduction of lattice strain will provide a crucial test to the
model proposed. This can be achieved, for example, by the
application of external pressure.\cite{yenchu:06}

The way pressure affects the magnetic and FE orders is different
from the action of an external magnetic field. The field couples
directly to the magnetic moments and affects the magnetic order in
aligning the moments with the field. This can stabilize or suppress
magnetic phases depending on the field orientation and the specifics
of the magnetic order as observed in recent
experiments.\cite{lawes:05} The application of external pressure,
however, results in a change of the magnetic exchange coupling
constants by compressing the lattice, decreasing the interatomic
distances, and changing the bond angles between different ions. This
in turn will affect the magnetically ordered phases as well as the
FE displacements. A dramatic effect of hydrostatic pressure on the
ferroelectricity was recently revealed in multiferroic $RMn_2O_5$
manganites.\cite{delacruz:06b} We have therefore investigated the
effect of hydrostatic pressure on the LTI phase of $Ni_3V_2O_8$ by
measuring the FE polarization under high-pressure conditions. The
results are correlated with strong lattice anomalies observed at the
LTI $\Rightarrow$ C phase boundary in thermal expansion
measurements.

Single crystals of $Ni_3V_2O_8$ (7 mm diameter and 50 mm long) have
been grown using the floating zone furnace. For measuring the
thermal expansivities along all three crystallographic orientations
samples of typical edge length of 2 to 3 mm were cut from the single
crystal and mounted in a high-resolution capacitance
dilatometer.\cite{delacruz:05} The FE polarization was measured by
employing the pyroelectric current method. The samples were placed
in a beryllium-copper clamp cell that allowed for generation of
hydrostatic pressures up to 2 GPa.\cite{chu:74} The pressure was
measured in situ at low temperatures by monitoring the pressure
shift of the superconducting transition temperature of high purity
lead.\cite{smith:67} The pyroelectric current signal was measured
between 2 K and 10 K for a plate-like sample oriented along the
$b$-axis (the axis of the FE polarization) upon cooling and heating
with a small poling voltage applied. The FE polarization was
determined by integrating the pyroelectric current. The dielectric
constant of the same sample was measured using the HP4285A LCZ meter
at a frequency of 100 kHz.

At ambient pressure the FE polarization $P$ arises at the phase
transition from the HTI phase (sinusoidal spin density wave) to the
LTI phase (helical spin modulation) at $T_{C1}$=6.5 K and it drops
to zero at the transition into the commensurate magnetic phase
($T_{C2}$=3.9 K). The measured $P(T)$ as shown in the right inset to
Fig. 1 is comparable with previous reports.\cite{lawes:05} The
dielectric constant $\varepsilon$ shown in Fig. 1 exhibits a sharp
peak at $T_{C1}$ with a small temperature hysteresis ($<$ 0.03 K)
and a small but distinct step at the first order phase transition at
$T_{C2}$ (left inset of Fig. 1). The transition from the
paramagnetic (PM) phase into the HTI phase is also detected in
$\varepsilon(T)$ in form of a sudden change of slope at $T_N$=9.8 K.
Under hydrostatic pressure the peak of $\varepsilon$ shown in Fig.
2b slightly shifts to lower temperatures, however, the overall shift
at 2 GPa is very small, $\triangle T$=$-$0.35 K. It is remarkable
that the peak width increases dramatically above 1.6 GPa. The LTI to
HTI phase boundary is well defined by the peak of $\varepsilon(T)$.
The stability of the ferroelectric LTI phase towards low
temperature, however, is difficult to extract from the pressure
dependence of the small step-like feature of $\varepsilon(T)$ at
$T_{C2}$ shown in the inset of Fig. 1. At low pressure $T_{C2}$
increases but at higher pressure the $\varepsilon$-step moves into
the tail of the large $\varepsilon(T)$-peak and cannot be resolved
any further.

Measurements of the FE polarization is the definitive signature of
the existence of ferroelectricity and thus provide an unambiguous
way to determine the stability range of the FE LTI phase. Fig. 2a
shows the FE polarization $P$ at different pressures. The
low-temperature drop of $P$ at $T_{C2}$ shifts to higher temperature
in accordance with the dielectric constant measurements. At the same
time the magnitude of $P$ is dramatically reduced and the
spontaneous polarization no longer exists at pressures above 1.64
GPa. At this critical pressure the FE state is completely
suppressed. The onset of ferroelectricity indicated by the rise of
the polarization at $T_{C1}$ is consistent with the pressure
dependence of the sharp peak of the dielectric constant (Fig. 2b).
The peak height of $\varepsilon$ decreases quickly in analogy with
the suppression of the ferroelectric polarization. The
pressure-temperature phase diagram for $Ni_3V_2O_8$ shown in Fig. 3
is constructed from the high pressure measurements of both the
dielectric constant and the ferroelectric polarization. The phase
boundary between the PM and HTI phases was determined from the
$\varepsilon(T)$ anomaly at $T_N$ (shown in Fig. 1, left inset).
$T_N$ slightly increases under pressure whereas $T_{C1}$ decreases.
$T_{C2}$ increases with pressure and merges with $T_{C1}$ at a
tricritical point at 1.64 GPa and 6.2 K. The ferroelectric LTI phase
ends at the tricritical point and above the critical pressure the
transition from the incommensurate HTI phase proceeds directly into
the commensurate C phase.

The $Ni_3V_2O_8$ compound displays a layered structure and thus is
expected to experience an anisotropic strain under a hydrostatic
pressure. For a detailed analysis of the pressure effect on the
compound, data of compressibilitiy along different crystal axes is
required. While compressibility data for $Ni_3V_2O_8$ are not
available, measurements of the anisotropy of the thermal expansion
and the lattice anomalies at the phase transitions can be utilized
to characterize the anisotropic properties of the structure. The
thermal expansivities along the principal crystallographic
orientations have been measured and the data are displayed in Fig.
4. Whereas small anomalies of the lattice parameters are barely
detected at $T_N$ and $T_{C1}$, the transition from the LTI phase
into the C phase at $T_{C2}$ exhibits the largest anomalies in form
of sizable abrupt changes of $a$, $b$, and $c$. The anisotropy of
these anomalies ($a$, $b$ expand but $c$ contracts upon cooling
through $T_{C2}$) shows the strongly anisotropic character of the
spin-lattice coupling at this transition. Note that the change of
the volume $V$ is relatively small though the uniaxial strain along
the crystallographic orientations is large. The relative changes of
$a$, $b$, $c$, and $V$ upon cooling through $T_{C2}$ as estimated
from the dilatometric measurements are $\Delta a/a=1.65*10^{-5}$,
$\Delta b/b=3.97*10^{-5}$, $\Delta c/c=-8.24*10^{-5}$, and $\Delta
V/V=-2.62*10^{-5}$, respectively. The volume of the low-temperature
C-phase is smaller than the volume of the LTI-phase which could
explain the increase of $T_{C2}$ since pressure will stabilize the
smaller volume phase. However, the effects of pressure on the
microscopic magnetic exchange constants in the anisotropic structure
of $Ni_3V_2O_8$ should be essential as well.

A comprehensive investigation of the magnetic orders by Kenzelmann
et al.\cite{kenzelmann:06} indicates that nearest (NN) and
next-nearest (NNN) neighbor superexchange interactions (parameters
$J_1$ and $J_2$, respectively), as well as the single ion anisotropy
(SA) (parameter $K$) of the nickel ions at the spine sites of the
Kagom\'{e} lattice have to be considered as a minimum to understand
the sequence of phase transitions from PM $\Rightarrow$ HTI
$\Rightarrow$ LTI $\Rightarrow$ C upon decreasing temperature. A
qualitative phase diagram could be constructed by evaluating a one
dimensional model in mean field approximation including the effects
of NN, NNN, and SA. For a ratio of $J_1/J_2$=2.56, the solution of
the model calculation reproduces the IC magnetic orders with the
experimentally observed wave vector of
$\overrightarrow{q}$=(0.27,0,0). The simplified model does describe
the observed sequence of phase transitions if the remaining free
parameter $K/J_1$ is chosen appropriately as
$K/J_1$=0.6.\cite{kenzelmann:06} It is interesting that the model
phase diagram, $T/J_1$ versus $K/J_1$, exhibits very similar
features as our pressure-temperature phase diagram of Fig. 3. With
increasing ratio $K/J_1$ the Ne\'{e}l transition temperature $T_N$
as well as $T_{C2}$ increase while $T_{C1}$ decreases until it
merges with $T_{C2}$ resulting in a tricritical point. According to
our high-pressure phase diagram, the external pressure affects all
three critical temperatures in exactly the same way as $K/J_1$ does
in the model phase diagram. It therefore appears conceivable that
pressure can be considered to increase the model parameter $K/J_1$.

The suppression of ferroelectricity by pressure in $Ni_3V_2O_8$ is
in contrast to the observation that pressure increases the FE
polarization at low temperatures in $HoMn_2O_5$, another
multiferroic compound with ferroelectricity induced by frustrated
magnetic orders.\cite{delacruz:06,delacruz:06b} The major difference
between the FE phases in $Ni_3V_2O_8$ and $HoMn_2O_5$ is the
commensurability of the magnetic phase associated with the
ferroelectricity. The magnetic order in the FE phase in $Ni_3V_2O_8$
is incommensurate whereas it is commensurate in the case of
$HoMn_2O_5$. It appears that external pressure always favors the
commensurate magnetic orders which can explain the opposite pressure
effects in the two multiferroic compounds. In the case of
$Ni_3V_2O_8$ the pressure effect can be related to the single ion
anisotropy (i.e. the parameter $K/J_1$) in reference to the simple
model discussed before.\cite{kenzelmann:06} In $HoMn_2O_5$ as well
as in other rare earth $RMn_2O_5$, however, the role of the magnetic
anisotropy in relation to the different magnetic exchange couplings
among the $Mn^{3+}$-, $Mn^{4+}$-, and $R^{3+}$-moments is not yet
clear.\cite{blake:05}

Although the comparison of the model phase
diagram\cite{kenzelmann:06} and the pressure-temperature diagram of
this work shows striking similarities, a more detailed experimental
as well as theoretical investigation is needed to clarify the effect
of pressure on the microscopic exchange coupling and anisotropy
constants. A more sophisticated model should take into account the
specifics of the 2D Kagom\'{e} staircase structure which is a
frustrated system by geometry. The interactions between different
planes along the $b$-axis may also play an essential role in
stabilizing long-range magnetic and FE orders. Some details have
been discussed extensively by Kenzelmann et al.\cite{kenzelmann:06}
and it was found that, besides the NN, NNN, and SA interactions
discussed above, much weaker Dzyaloshinskii-Moriya as well as
pseudo-dipolar interactions contribute to the complex magnetic
orders of the Ni-spine spins and their coupling to the Ni moments at
cross-tie positions. An important step towards a microscopic theory
of the magnetoelectric effect in $Ni_3V_2O_8$ is the recent work of
Harris et al.\cite{harris:06} Based on inelastic neutron scattering
experiments and a first-principles calculation of phonon modes, the
most relevant phonons that may give rise to the magnetoelectric
coupling and the FE distortions have been identified. The lattice
distortions at the FE transitions derived from this investigation
are so small that they are hardly observed in neutron scattering
experiments. It is the extraordinary high resolution of our
dilatometric measurements that allows to detect the lattice strain
at the magnetic and FE phase transitions in multiferroic
materials.\cite{delacruz:06} However, in $Ni_3V_2O_8$, a sizable
strain is only seen at the low-$T$ lock-in transition into the
commensurate phase, as shown in Fig. 4. At the onset of
ferroelectricity at $T_{C1}$ an anomaly of the lattice parameters is
barely detectable. It is the symmetry of the magnetic structure and
of the associated small non-centrosymmetric displacement of the ions
that is essential at this transition resulting in a net FE
polarization below $T_{C1}$. The expansion data prove the strong
anisotropy of the structure and the magnetically ordered phases. Any
forthcoming theory has to account for the anisotropic character of
the system. Experimentally, the anisotropy could be controlled by
applying uniaxial pressure along the main crystallographic
directions. This would, together with first-principle calculations
of the strain dependence of the most important magnetic exchange and
anisotropy parameters, facilitate a better understanding of the
strain effects on the magnetoelectric coupling in $Ni_3V_2O_8$.

The pressure-temperature phase diagram of Fig. 3 implies that in a
narrow temperature range pressure will change the commensurability
of the magnetic order as well as the detailed spin alignment in
crossing the $T_{C2}$ phase boundary. This prediction should be
confirmed by the results of neutron scattering experiments conducted
under high-pressure conditions. While the wave vector of the
magnetic modulation is sensitive to the ratio of the NN and NNN
exchange integrals\cite{kenzelmann:06} high-pressure neutron
scattering is the ideal tool to further investigate the pressure
effect on the magnetic coupling parameters. It can be expected that
even relatively low pressure does change the characteristic wave
vector of the magnetic orders in the HTI and LTI phases. Since big
single crystals of $Ni_3V_2O_8$ can be grown in a floating zone
furnace these experiments are feasible and they may provide the
final key to solve the complex problem of magnetic and ferroelectric
orders and their mutual interactions in the multiferroic
$Ni_3V_2O_8$.

\begin{acknowledgments}
This work is supported in part by the T.L.L. Temple Foundation, the
J. J. and R. Moores Endowment, and the State of Texas through TCSUH
and at LBNL through the US DOE, Contract No. DE-AC03-76SF00098.
\end{acknowledgments}

\bibliographystyle{phpf}


\newpage

\begin{figure}
\caption{Temperature dependence of the dielectric constant of
$Ni_3V_2O_8$ near the FE transitions. Full circles: cooling, Open
circles: heating. Left inset: Step of $\varepsilon$ at the lock-in
transition into the commensurate phase. Right inset: FE
polarization.}
\caption{(Color online) Temperature dependence of (a) the FE
polarization and (b) the dielectric constant at different pressures.
Different $\varepsilon(T)$-curves are vertically offset for better
clarity.}
\caption{Pressure-temperature phase diagram of $Ni_3V_2O_8$.}
\caption{(Color online) Lattice strain along $a$-, $b$-, and
$c$-axes below 10 K. The reference length $L_0$ is the lattice
parameter at 10 K.}
\end{figure}


\begin{thebibliography}{10}

\bibitem{lawes:04}
G.~Lawes, M.~Kenzelmann, N.~Rogado, K.~H. Kim, G.~A. Jorge, R.~J.
Cava,
  A.~Aharony, O.~Entin-Wohlman, A.~B. Harris, T.~Yildirim, Q.~Z. Huang,
  S.~Park, C.~Broholm, and A.~P. Ramirez,
\newblock Phys. Rev. Lett. {\bf 93}, 247201 (2004).

\bibitem{lawes:05}
G.~Lawes, A.~B. Harris, T.~Kimura, N.~Rogado, R.~J. Cava,
A.~Aharony,
  O.~Entin-Wohlman, T.~Yildirim, M.~Kenzelmann, C.~Broholm, and A.~P. Ramirez,
\newblock Phys. Rev. Lett. {\bf 95}, 087205 (2005).

\bibitem{kenzelmann:06}
M.~Kenzelmann, A.~B. Harris, A.~Aharony, O.~Entin-Wohlman,
T.~Yildirim,
  Q.~Huang, S.~Park, G.~Lawes, C.~Broholm, N.~Rogado, R.~J. Cava, K.~H. Kim,
  G.~Jorge, and A.~P. Ramirez,
\newblock Phys. Rev. B {\bf 74}, 014429 (2006).

\bibitem{harris:06}
A.~B. Harris, T.~Yildirim, A.~Aharony, and O.~Entin-Wohlman,
\newblock Phys. Rev. B {\bf 73}, 184433 (2006).

\bibitem{harris:06b}
A.~B. Harris,
\newblock J. Appl. Phys. {\bf 99}, 08E303 (2006).

\bibitem{wilson:06}
N.~R. Wilson, O.~A. Petrenko, and G.~Balakrishnan,
\newblock cond-mat/0610123, unpublished.

\bibitem{mostovoy:06}
M.~Mostovoy,
\newblock Phys. Rev. Lett. {\bf 96}, 067601 (2006).

\bibitem{yenchu:06}
F. Yen, unpublished. Preliminary results presented at the APS March
Meeting,
  2006 (C. W. Chu, Bull. Am. Phys. Soc. 51, p. 52, 2006).

\bibitem{delacruz:06b}
C.~R. dela Cruz, B.~Lorenz, M.~M. Gospodinov, and C.~W. Chu,
\newblock J. Mag. Mag. Mat.  (2006),
\newblock in press.

\bibitem{delacruz:05}
C.~R. dela Cruz, F.~Yen, B.~Lorenz, Y.~Q. Wang, Y.~Y. Sun, M.~M.
Gospodinov,
  and C.~W. Chu,
\newblock Phys. Rev. B {\bf 71}, 060407(R) (2005).

\bibitem{chu:74}
C.~W. Chu and L.~R. Testardi,
\newblock Phys. Rev. Lett. {\bf 32}, 766 (1974).

\bibitem{smith:67}
T.~F. Smith and C.~W. Chu,
\newblock Phys. Rev. {\bf 159}, 353 (1967).

\bibitem{delacruz:06}
C.~R. dela Cruz, F.~Yen, B.~Lorenz, M.~M. Gospodinov, C.~W. Chu,
W.~Ratcliff,
  J.~W. Lynn, S.~Park, and S.-W. Cheong,
\newblock Phys. Rev. B {\bf 73}, 100406(R) (2006).

\bibitem{blake:05}
G.~R. Blake, L.~C. Chapon, P.~G. Radaelli, S.~Park, N.~Hur, S.-W.
Cheong, and
  J.~Rodriguez-Carvajal,
\newblock Phys. Rev. B {\bf 71}, 214402 (2005).

\end{thebibliography}

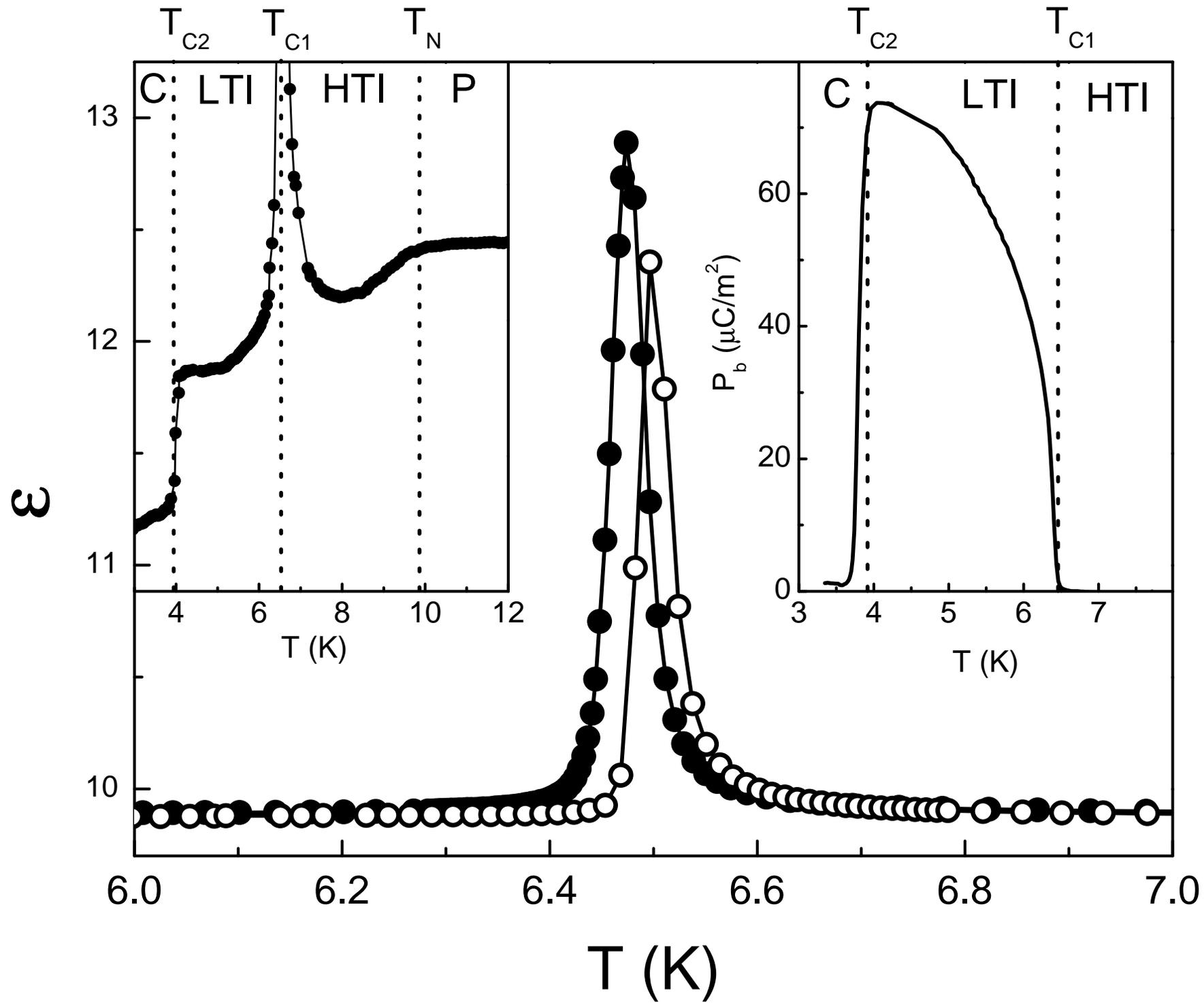
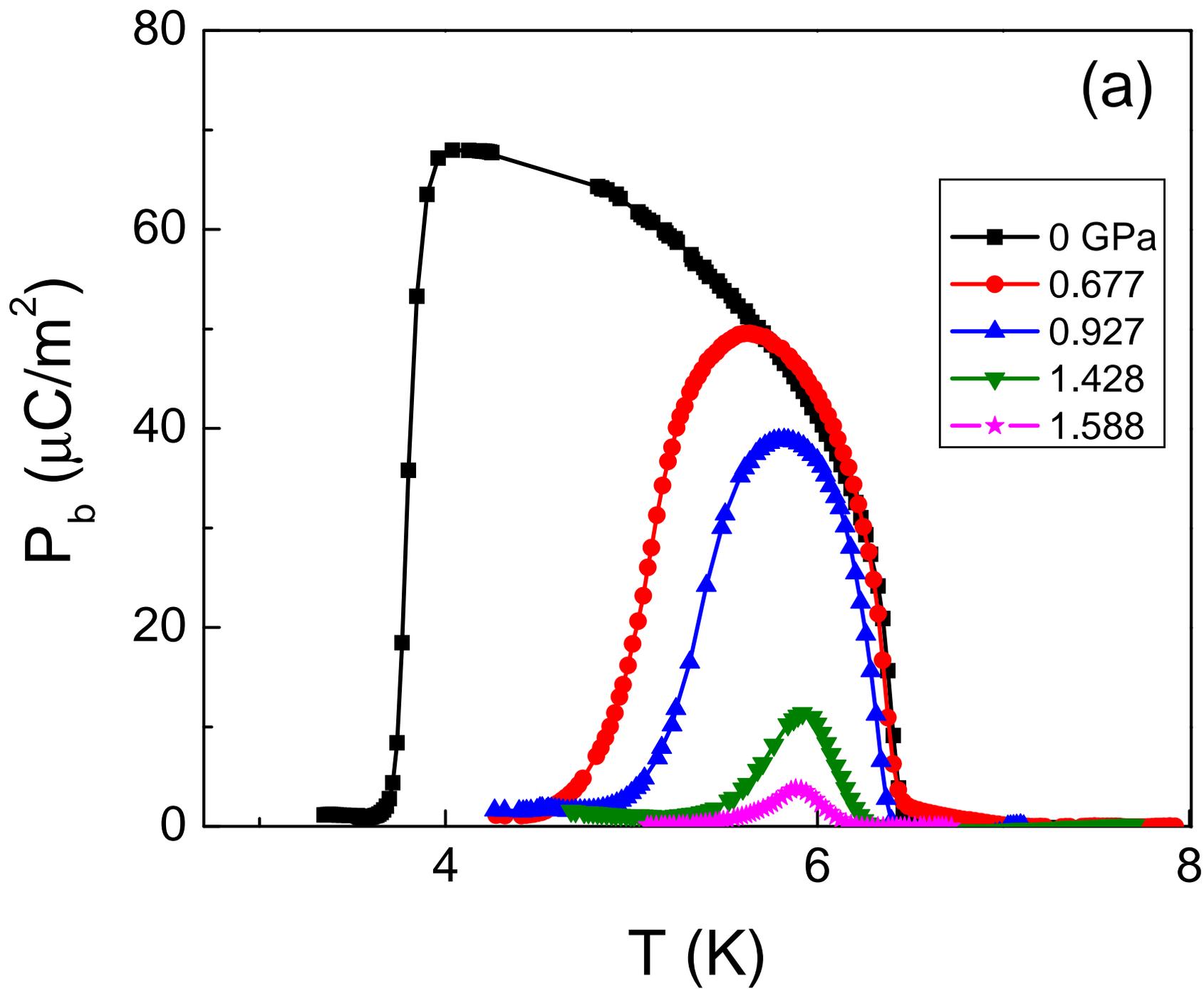
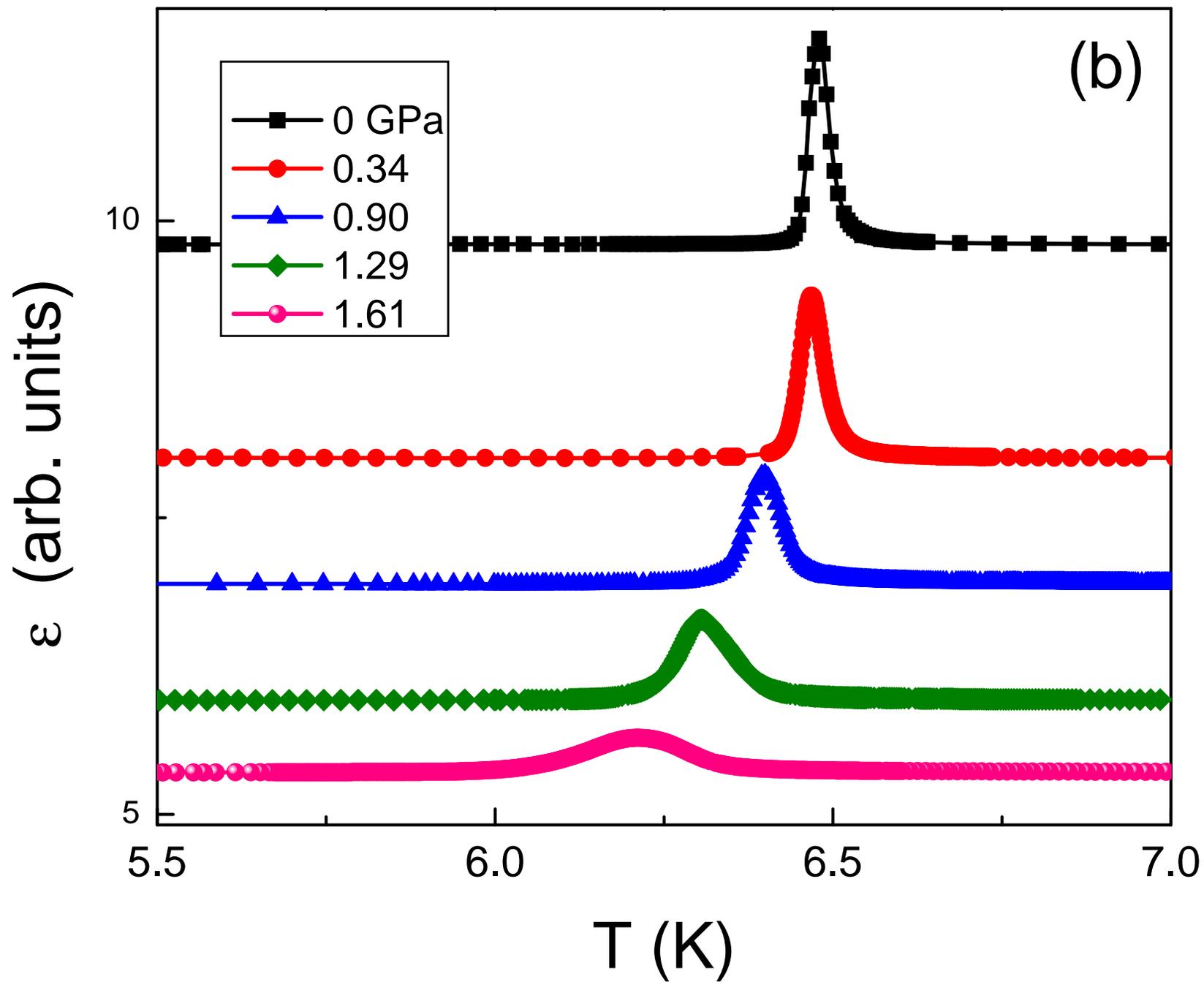
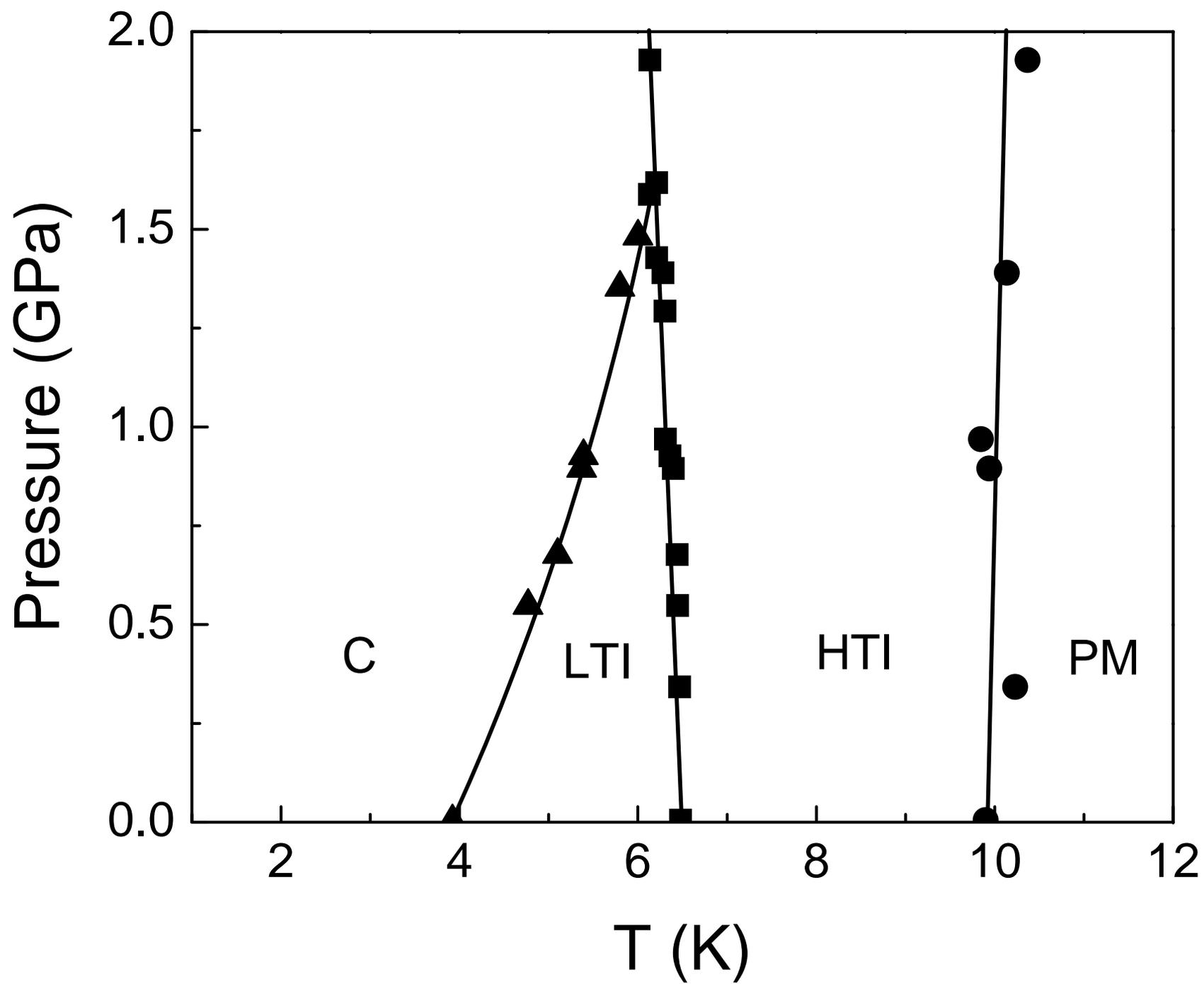
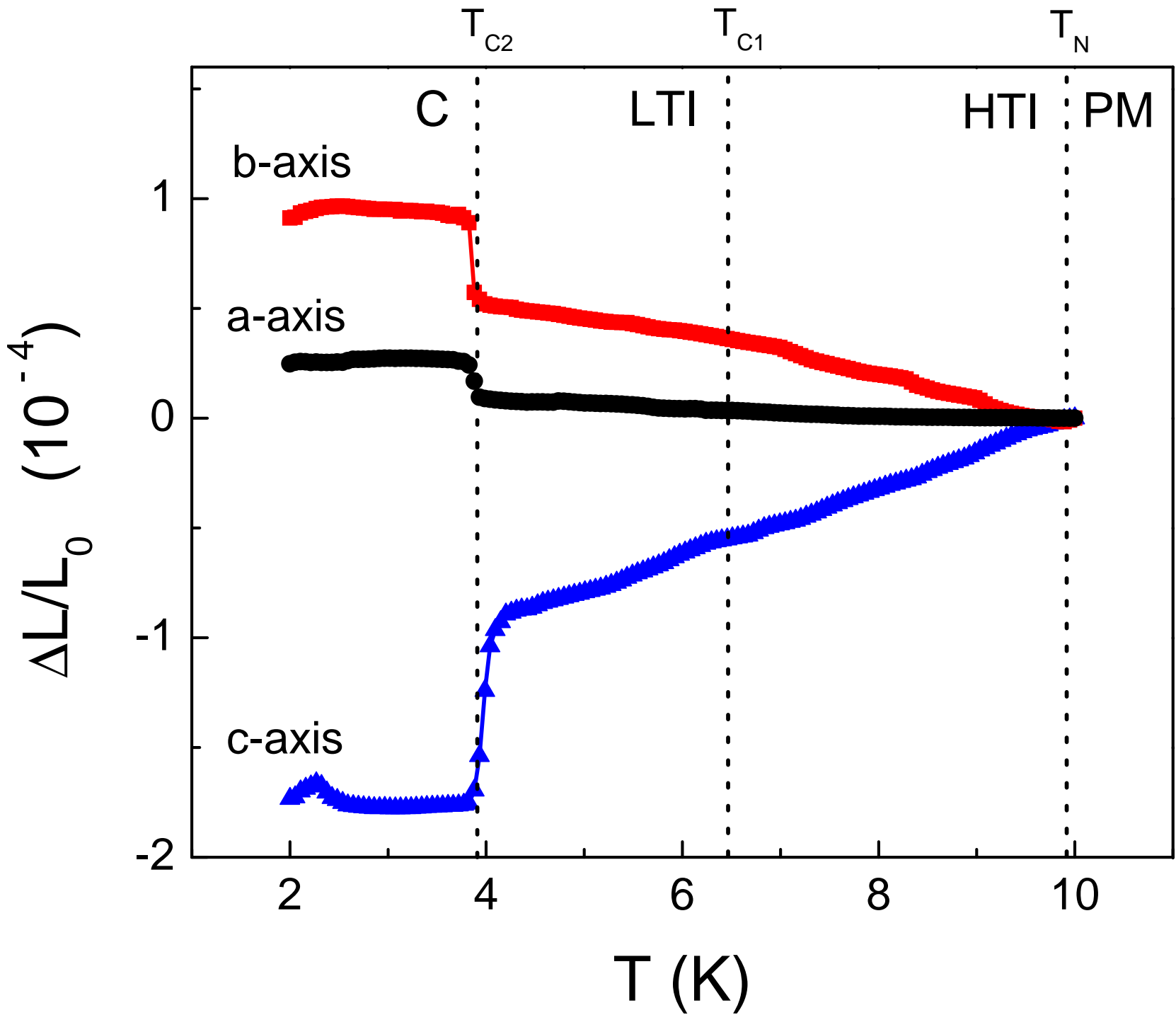
\end{document}